\def\year{2024}\relax
\documentclass[letterpaper]{article} 
\usepackage{aaai22}  
\usepackage{times}  
\usepackage{helvet}  
\usepackage{courier}  
\usepackage[hyphens]{url}  
\usepackage{graphicx} 
\usepackage{natbib}  
\usepackage{caption} 
\DeclareCaptionStyle{ruled}{labelfont=normalfont,labelsep=colon,strut=off} 
\usepackage{algorithm}
\usepackage{algorithmic}
\usepackage{subcaption}
\usepackage{multirow}
\usepackage{multicol}

\usepackage{xcolor}
\newcommand{\answerYes}[1]{\textcolor{blue}{#1}} 
\newcommand{\answerNo}[1]{\textcolor{teal}{#1}} 
\newcommand{\answerNA}[1]{\textcolor{gray}{#1}} 
 
\usepackage{newfloat}
\usepackage{listings}
\floatstyle{ruled}
\newfloat{listing}{tb}{lst}{}
\floatname{listing}{Listing}
\title{Language-Agnostic Modeling of Wikipedia Articles for Content Quality Assessment across Languages}
\author {
    Paramita Das\textsuperscript{\rm 1},
    Isaac Johnson\textsuperscript{\rm 2},
    Diego Saez-Trumper\textsuperscript{\rm 2},
    Pablo Aragón\textsuperscript{\rm 2}
}
\affiliations {
    \textsuperscript{\rm 1} IIT Kharagpur, India\\
    \textsuperscript{\rm 2} Wikimedia Foundation\\
    paramita.das@iitkgp.ac.in, isaac@wikimedia.org, diego@wikimedia.org, paragon@wikimedia.org
}

\begin{document}

\maketitle

\begin{abstract}
Wikipedia is the largest web repository of free knowledge. Volunteer editors devote time and effort to creating and expanding articles in more than 300 language editions. As content quality varies from article to article, editors also spend substantial time rating articles with specific criteria. However, keeping these assessments complete and up-to-date is largely impossible given the ever-changing nature of Wikipedia. To overcome this limitation, we propose a novel computational framework for modeling the quality of Wikipedia articles. 

State-of-the-art approaches to model Wikipedia article quality have leveraged machine learning techniques with language-specific features. In contrast, our framework is based on language-agnostic structural features extracted from the articles, a set of universal weights, and a language version-specific normalization criterion. Therefore, we ensure that all language editions of Wikipedia can benefit from our framework, even those that do not have their own quality assessment scheme. Using this framework, we have built datasets with the feature values and quality scores of all revisions of all articles in the existing language versions of Wikipedia. We provide a descriptive analysis of these resources and a benchmark of our framework. In addition, we discuss possible downstream tasks to be addressed with these datasets, which are released for public use.
\end{abstract}

\section{Introduction} 

Wikipedia is not only one of the most popular websites but also one of the largest free knowledge repositories in the world. Millions of people access Wikipedia daily in search of information on a multitude of topics~\cite{10.1145/3038912.3052716}. Furthermore, several search engines and AI-powered services rely on data extracted from Wikipedia articles~\cite{mikolov2017advances}. As a consequence, the quality of the articles has a key impact in the age of the knowledge society.

Wikipedia articles range in quality from rich, well-illustrated, fully-referenced articles that fully cover their topic and are easy to read to single sentence stubs that define the topic of the article but do not offer much more information. 
Editors have developed rich rubrics for how to evaluate the quality of Wikipedia articles and are constantly assessing article quality to assist in coordinating work on the wikis. However, Wikipedia is ever-changing though, which makes it time-consuming (and largely impossible) for editors to keep these quality assessments complete and up-to-date. Although several automatic quality models have been proposed~\cite{hasan2009automatic, warncke2013tell,dang2016measuring,bassani2019automatically}, most of them are language-specific, finely-tuned to the dynamics and existing quality classes of a particular language edition. Some other approaches have explored machine learning models for multilingual assessment of the quality of Wikipedia articles~\cite{halfaker2020ores}. Nevertheless, these models rely on content-dependent features and require substantial efforts to generate training datasets, as the criteria for assessing article quality are not consistent across all language versions of Wikipedia. 

To overcome these limitations when modeling the quality of Wikipedia articles across different languages, we present a framework based on language-agnostic features. We extend previous approaches~\cite{warncke2013tell,lewoniewski2019multilingual} by proposing 6 different features that we extract from the implicit structure of Wikitext - the markup language used to write Wikipedia. Our framework relies on a heuristic approach combining universal feature weights and a normalization criterion derived from each language version of Wikipedia in order to assess the quality of articles across different languages. We apply the framework on the full dump of revisions from all language versions of Wikipedia to show how these language-agnostic features are able to capture article quality. Furthermore, we evaluate our model and benchmark it against two different baselines in order to discuss the implications of language-agnostic modeling.

Scientists from diverse research disciplines 
have found Wikipedia to be a key repository of free available knowledge. However, resources that are made available are often related to a very selected subgroup of popular language versions among the over 300 existing ones~\cite{johnson2022considerations}. For that reason, our language-agnostic modeling framework has been created as a resource to provide knowledge in all languages, following an inspirational principle of knowledge equity~\cite{sefidari202120}. All the data generated from this work is already available\footnote{\url{https://doi.org/10.5281/zenodo.10495081}}. Therefore, we expect a diversity of research communities to benefit from this work.

The remainder of this paper is organized as follows. We review related work on Wikipedia focused on language-agnostic approaches and article quality assessment in the next section. Then, we present our set of language agnostic-features, the feature extraction process, and the resulting dataset of features from over 2 billion revisions. Our framework of quality modeling of Wikipedia articles with these features is described in the following section, including a model evaluation and benchmarking. Finally, we conclude in the last section by discussing possible downstream tasks to be addressed with our datasets, together with ethical and FAIR considerations.
\section{Related Work}\label{sec:related}

Wikipedia has proven to be a fruitful source for research datasets~\cite{flock2017toktrack,miquel2019wikipedia,consonni2019wikilinkgraphs,mitrevski2020wikihist,valentim2021tracking,meier2022twikil}. To provide the context of our framework and datasets and to position our contributions to the state of the art, we review related work on language-agnostic approaches to characterize Wikipedia content and on quality assessment of Wikipedia articles.

\subsection{Language-Agnostic Approaches}

Recent contributions have leveraged knowledge from English Wikipedia for diverse tasks, e.g., providing explainable search results~\cite{10.1145/3477495.3532067}, evidence retrieval for fact verification~\cite{10.1145/3477495.3531827}, text stance detection~\cite{10.1145/3477495.3531807}, sentence retrieval for open-ended dialogues~\cite{10.1145/3477495.3531727}, question answering~\cite{10.1145/3477495.3531734,10.1145/3477495.3531753}, 
named entities and relationships retrieval from articles~\cite{10.1145/3477495.3536322,10.1145/3477495.3531742}, etc. However, one of the most important aspects of Wikipedia is its multilingual character. Wikipedia has been referred to as the ``Web 2.0 Tower of Babel''~\cite{10.1145/1753326.1753370} in which knowledge is created and maintained by volunteers in more than 300 language versions, each independently. Therefore, some resources that have been made available to support information retrieval tasks in multiple languages also rely on Wikipedia~\cite{10.1145/3477495.3531731,10.1145/3404835.3463257,10.1145/3477495.3531662,10.1145/3477495.3531886}.

Multilingual approaches represent an important advance as many communities and contexts can benefit from NLP resources. However, these approaches are often limited by the availability of language-specific data, resulting in an uneven representation of the plethora of existing languages. On the one hand, one way of addressing scarcity in low-resourced languages is machine translation. In fact, we are witnessing the rise of initiatives bases on participatory approaches to make NLP research with Wikipedia low-resourced languages more scalable~\cite{nekoto2020participatory}. On the other hand, there is a growing trend towards research on Wikipedia through language-agnostic approaches. While the performance of models that exclude language-dependent features may be limited to a certain degree, their applicability to any language is almost straightforward. For this reason, several language-agnostic approaches have been proposed in recent years for different tasks on Wikipedia, e.g., spam users identification~\cite{Green_Spezzano_2017}, topic classification of articles~\cite{johnson2021language}, cross-lingual topic modeling~\cite{10.1145/3442381.3449805}, knowledge propagation tracking across languages~\cite{valentim2021tracking}, and evaluation of gender biases~\cite{Beytia_Agarwal_Redi_Singh_2022}.

\subsection{Article Quality Assessment}

Maintaining the quality of content in parallel with its rapid evolution is a both crucial and overwhelming task in large peer-to-peer production systems such as Wikipedia. Due to the immense cost involved in the process of monitoring the quality of articles in Wikipedia, several works have proposed different algorithmic methods for automatic quality assessment~\cite{moas2023automatic}.
The most intuitive approach is to provide support with machine learning supervised models trained on the basis of domain-specific feature engineering~\cite{warncke2013tell,dang2016measuring,bassani2019automatically}. Authors of these works have suggested a wider set of qualitative features related to the content of Wikipedia articles and perform the classification using machine learning models. Other approaches have proposed a hybrid model combining neural network LSTM-based document embeddings and handcrafted features~\cite{shen2017hybrid}. Further, subsequent works have attempted end-to-end neural models with the integration of qualitative features to better represent the articles and improve the performance of the classifiers~\cite{zhang2018history,shen2019joint}. Authors in~\cite{guda2020nwqm} proposed a multimodal framework for the automatic quality prediction that leverages contextual representation of main article text obtained from bidirectional transformers followed by a conditional summarization and inclusion of talk pages as the metadata about an article. 
All of these approaches assume the quality as a static attribute but it becomes obsolete often with the update of content. A complementary direction has been explored by~\cite{zhang2020mining,das2022quality} in which authors add temporal dimension to the static measurement of article quality and characterize the quality of articles through a number of dynamic changes in the quality states spanning over time. 

The aforementioned models were built using training datasets with content from English Wikipedia articles. Therefore, these approaches are limited to predicting the quality of the articles from the English version. An important step forward in multilingual assessment of the quality of Wikipedia articles has been the Objective Revision Evaluation Service (ORES)\footnote{\url{https://ores.wikimedia.org/}}~\cite{halfaker2020ores}. ORES uses structural and readability features as the indicators of article quality and classifies articles using a set of machine learning models. These models are language-specific and trained in a participatory way by the language communities of Wikipedia. As a consequence, models exist for only a few of the more than 300 existing Wikipedia language versions.
\section{Language-Agnostic Framework}\label{sec:features}

In this section, we focus on the framework that we developed to evaluate the quality of Wikipedia articles in over 300 languages. Our framework adopts a language-agnostic approach to represent Wikipedia articles, enabling us to assess the quality, such as the comprehensiveness of knowledge or information contained within them.

\subsection{Wikipedia Quality Assessment Scheme}
Almost each language edition of Wikipedia typically implements its own system for assessing the quality of articles, assigning labels based on their overall quality– good, mediocre, or substandard. These hierarchical quality rankings are based on several key factors, including topic coverage, content organization, structural style, etc. For instance, in English Wikipedia, articles are ranked using quality classes\footnote{\url{https://en.wikipedia.org/wiki/Wikipedia:Content\_assessment}}, such as– FA, GA, B, C, START, and STUB. FA represents the highest quality level, indicating articles that are comprehensive and well-written, while STUB denotes the lowest quality with minimal meaningful content that requires improvement in the overall structure of the article. Similar quality divisions exist in other language editions, like the French Wikipedia, where labels such as AdQ, BA, A, B, BD, and ébauche are used, in which AdQ represents the highest-quality article and ébauche stands for lowest quality, similar to STUB in English Wikipedia. 

The task of assessing article quality is primarily carried out by Wikipedia editors who mark their evaluations on the talk pages associated with each article. Talk pages serve as discussion forums where editors can engage in conversations related to an article. The current quality of an article is often documented on its corresponding talk page. These quality assessments, recorded on talk pages, are then collected and stored in the form of logs or statistics by automated bots. Typically, lower-level quality classes are assigned by individual editors based on their evaluations. To attain the highest levels of quality, such as FA or GA in English Wikipedia, potential articles are nominated by editors and subsequently reviewed by either individuals or panels. A selected number of articles that meet the necessary criteria are then included in the list of Featured or Good articles. Wikipedia maintains separate lists dedicated to articles that have successfully fulfilled all the requirements during the review process. These lists are periodically updated by the Wikipedia community.

In our evaluation scheme, we map the quality scores predicted by our framework, which range from 0 to 1, to the quality classes used in the English and French Wikipedia. This mapping enables us to assess the effectiveness and performance of our framework, as discussed in subsequent sections.

\subsection{Language Agnostic Features}
Many potential features could be incorporated into article quality models. Some features should be relevant to article quality in the traditional sense of a reader’s experience (e.g., page length, number of images) while others might just relate to what sort of work is considered important in building a high-quality Wikipedia article (e.g., quality of categories) even if there is not a clear, direct impact on the reader experience. Features related to edit history (e.g., number of editors) are generally not considered.

Our approach is inspired by previous work providing a grounded approach to developing quality models~\cite{warncke2013tell}, that served to propose rankings of relative quality and popularity assessment in multiple language versions of Wikipedia~\cite{lewoniewski2019multilingual}. For our modeling framework of article quality in any given language version of Wikipedia, we have designed the following set of language-agnostic features:
\begin{itemize}
    
    \item \textbf{Page length:} Square-root-normalized number of characters that exist in the Wikitext of the given revision.
    This feature might be the simplest measure of article quality but one with a fair bit of signal. Challenges include that different languages require substantially different numbers of characters to express the same ideas and thus the meaning of page length is not consistent across wikis.
    
    \item \textbf{References:} Number of ref tags that exist per normalized page length.
    Good Wikipedia articles should be verifiable, which generally means well-referenced. Different wikis sometimes use specific templates for references that are separate from the more universal ref tags, which makes their extraction very challenging.
    
    \item \textbf{Sections:} Number of headings (levels 2 and 3 only) per normalized page length. 
    Wikipedia articles are generally broken into sections to provide structure to the content. While more sections generally may suggest higher quality, it can be difficult to assert the appropriate number of sections for an article of a given length.
    
    \item \textbf{Wikilinks:} Square-root-normalized of wikilinks per normalized page-length. 
    Linking to other Wikipedia articles generally helps readers explore content and is a good practice. Different wikis have different norms, however, about where and when it is appropriate to add links.
    
    \item \textbf{Categories:} Number of categories (raw count; no transformation). 
    Categories make finding and maintaining content easier.    
    
    \item \textbf{Media:} Number of media files (raw count; no transformation) -- e.g., image, video, or audio files. 
    Multimedia content enriches Wikipedia articles, though certain topics are much easier to illustrate than others and certain norms lead to very high numbers of ``images'' -- e.g., including flags next to people's names to illustrate their nationality as is common in sports-related articles.
\end{itemize}

There are generally two ways to represent a given feature: raw count (e.g., number of references) or proportion (e.g., number of references / page length). The first approach purely emphasizes ``more content is better'' but is simple to interpret. The second approach emphasizes more controlled growth -- e.g., if adding a new section, the article might be penalized if that section does not have references. In the most extreme case, all features are proportions and a well-cited but too short and incomplete article with an image could be considered just as high-quality as an article featured by the community. In practice, some form of raw page length is probably always included to account for longer articles generally being higher quality.

\begin{figure}[h!]
    \centering
    \includegraphics[width=0.95\linewidth]{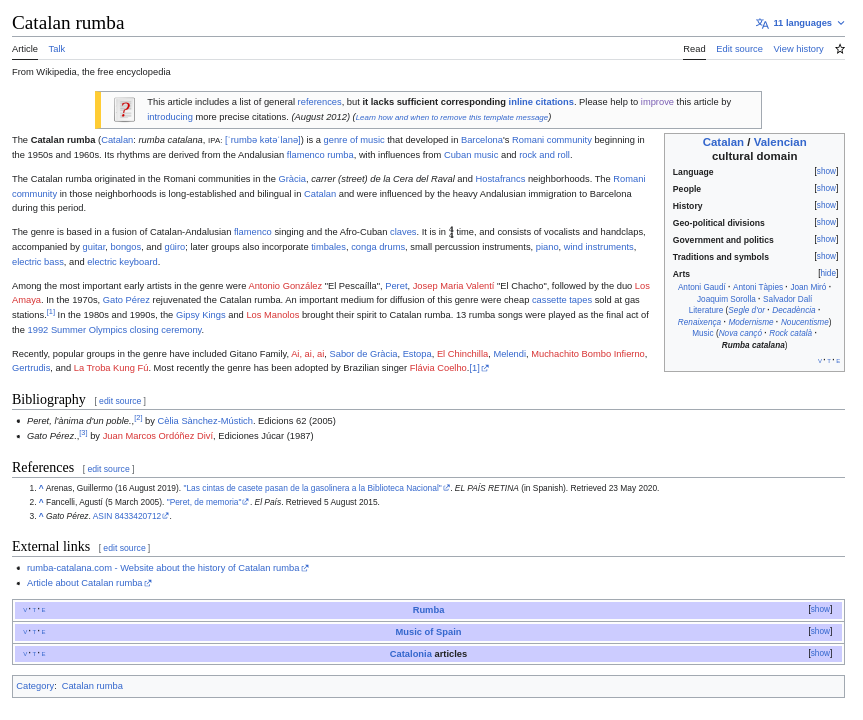}
    \caption{English Wikipedia article about \textit{Catalan Rumba}.}
    \label{fig:screenshot}
\end{figure}

Figure~\ref{fig:screenshot} shows a recent revision of the article of \textit{Catalan Rumba} in English Wikipedia (\textit{revision id 1171335521}) that has 3434~characters, 3~references, 3~sections, 41~wikilinks, 1~category, and 0~media files.

\subsection{Feature Extraction}

Wikipedia articles are not static, they evolve over time. Editors are responsible for both creating new articles and updating the existing ones by generating new versions, called \textit{revisions}. Revisions include the content of the corresponding version of the article in Wikitext format, as well as associated metadata such as the authoring editor, the timestamp or a descriptive comment by the editor about the revision.

The full history of revisions of Wikipedia articles is available in the XML dumps. To generate our dataset of language-agnostic features of Wikipedia articles, we first retrieve the Wikitext content of every revision of every article in every available language version of Wikipedia from the beginning to the end of 2022. It should be noted that we only consider pages that represent articles (i.e., main namespace\footnote{\url{https://en.wikipedia.org/wiki/Wikipedia:What_is_an_article}}) and that we omitted page redirects.
Then, we apply regular expressions to extract all the features in each revision. This could be also done with libraries for parsing Wikitext content such as \textit{mwparserfromhell}\footnote{\url{https://github.com/earwig/mwparserfromhell}}. However, we have found our approach with regular expressions on PySpark up to 10 times faster in medium-sized articles. 

\subsection{Dataset of Language-Agnostic Features}

With our feature extraction process, we generate a dataset of more than 2 billion revisions stored as CSV files (one file per language edition). Each row is a revision and the columns are the id of the revision (\textit{revision\_{}id}), the id of the page (\textit{page\_{}id}), and the values of the extracted language-agnostic features (\textit{page\_{}length}, \textit{num\_{}refs}, \textit{num\_{}sections}, \textit{num\_{}wikilinks}, \textit{num\_{}categories}, \textit{num\_{}media}).

\begin{figure*}[ht]
     \centering
     \begin{subfigure}[b]{0.33\textwidth}
         \centering
         \includegraphics[width=\textwidth]{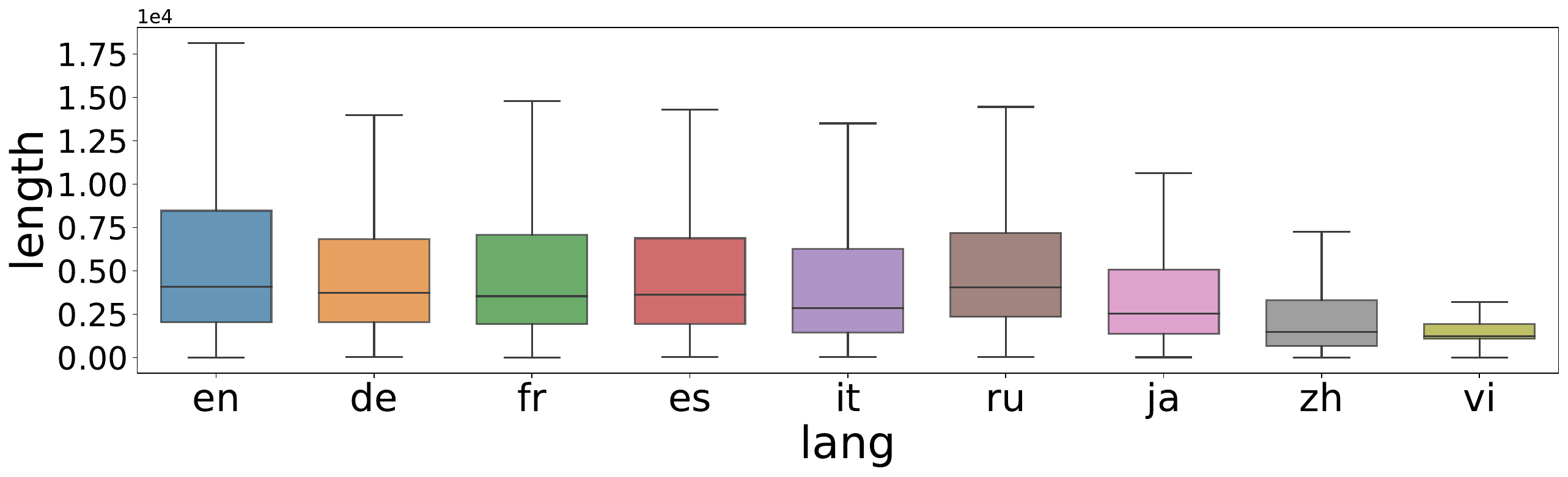}
         \caption{Page length.}
         \label{fig:page_length_boxplot}
     \end{subfigure}
     \hfill
     \begin{subfigure}[b]{0.33\textwidth}
         \centering
         \includegraphics[width=\textwidth]{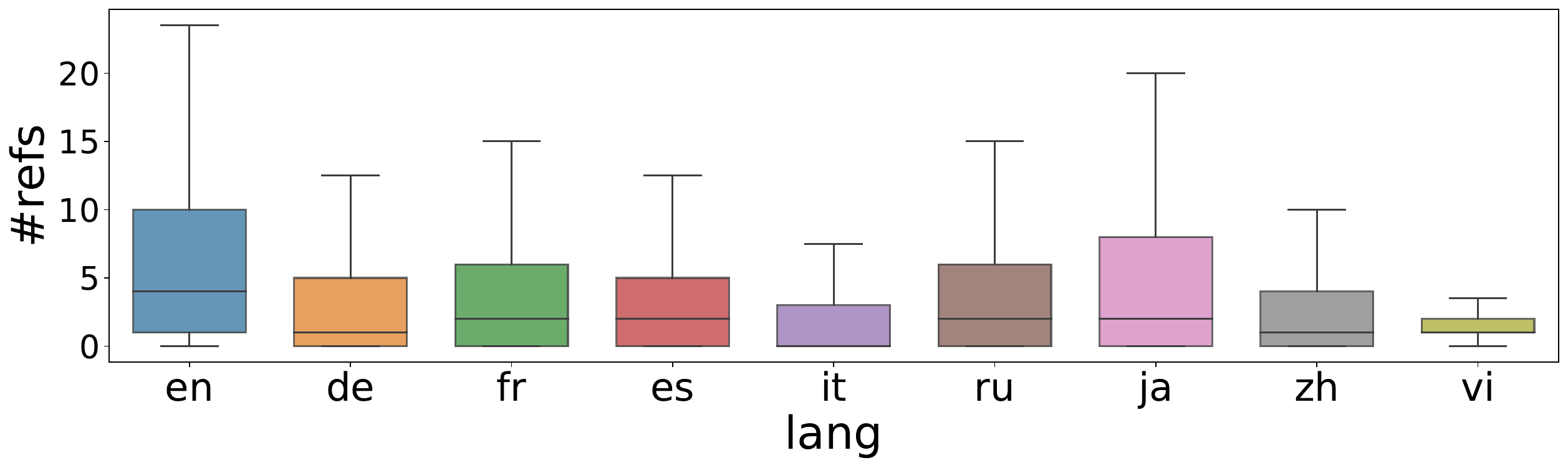}
         \caption{References.}
         \label{fig:num_refs_boxplot}
     \end{subfigure}
     \hfill
     \begin{subfigure}[b]{0.33\textwidth}
         \centering
         \includegraphics[width=\textwidth]{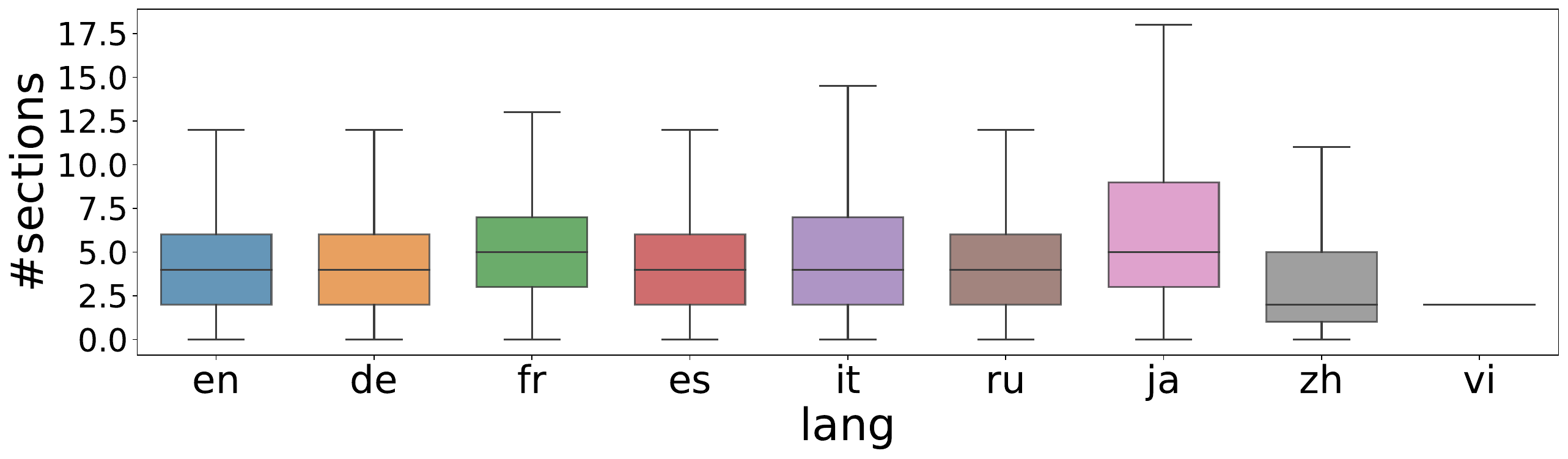}
         \caption{Sections.}
         \label{fig:num_headings_boxplot}
     \end{subfigure}       
     \hfill
     \begin{subfigure}[b]{0.33\textwidth}
         \centering
         \includegraphics[width=\textwidth]{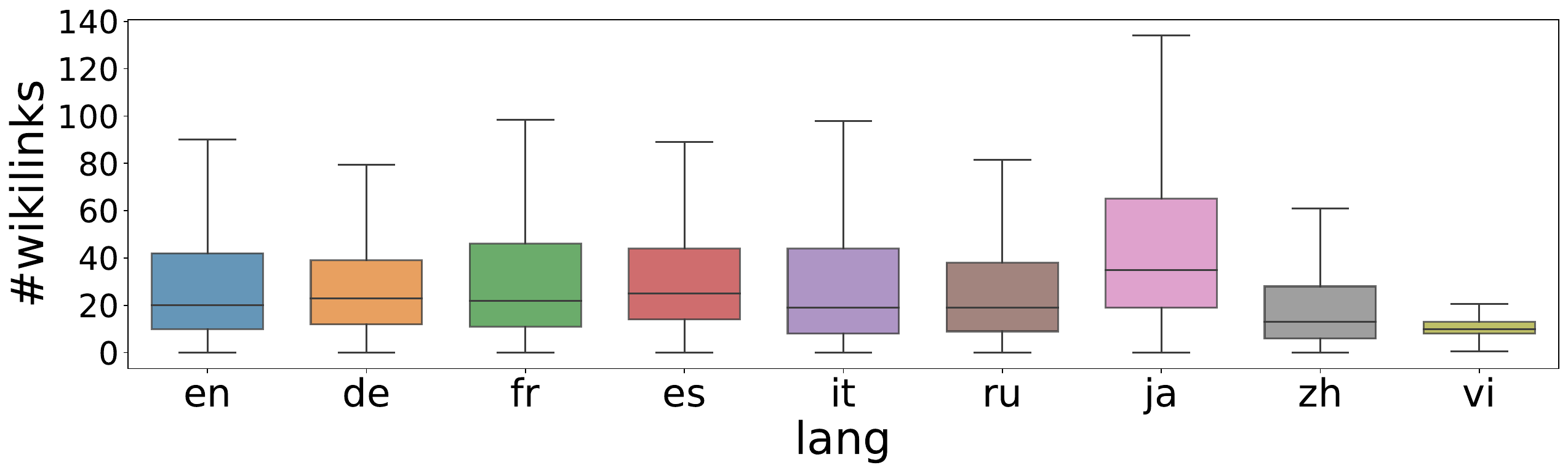}
         \caption{Wikilinks.}
         \label{fig:num_wikilinks_boxplot}
     \end{subfigure}
     \begin{subfigure}[b]{0.33\textwidth}
         \centering
         \includegraphics[width=\textwidth, trim={0 0 0 -2cm}, clip]{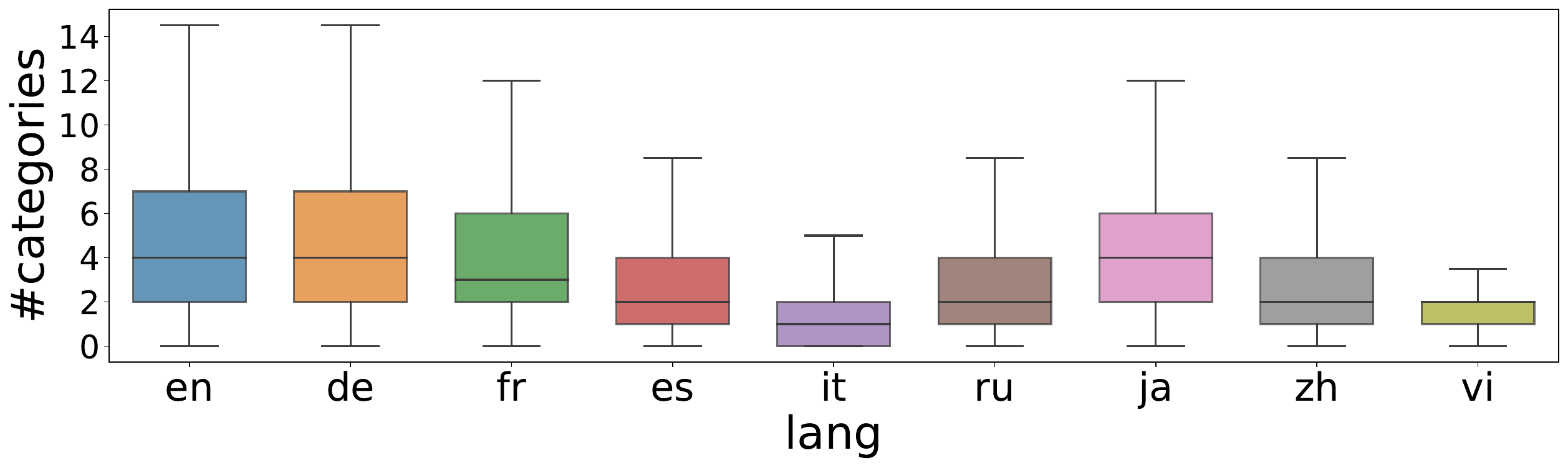}
         \caption{Categories.}
         \label{fig:num_categories_boxplot}
     \end{subfigure}
     \hfill
     \begin{subfigure}[b]{0.33\textwidth}
         \centering
         \includegraphics[width=\textwidth]{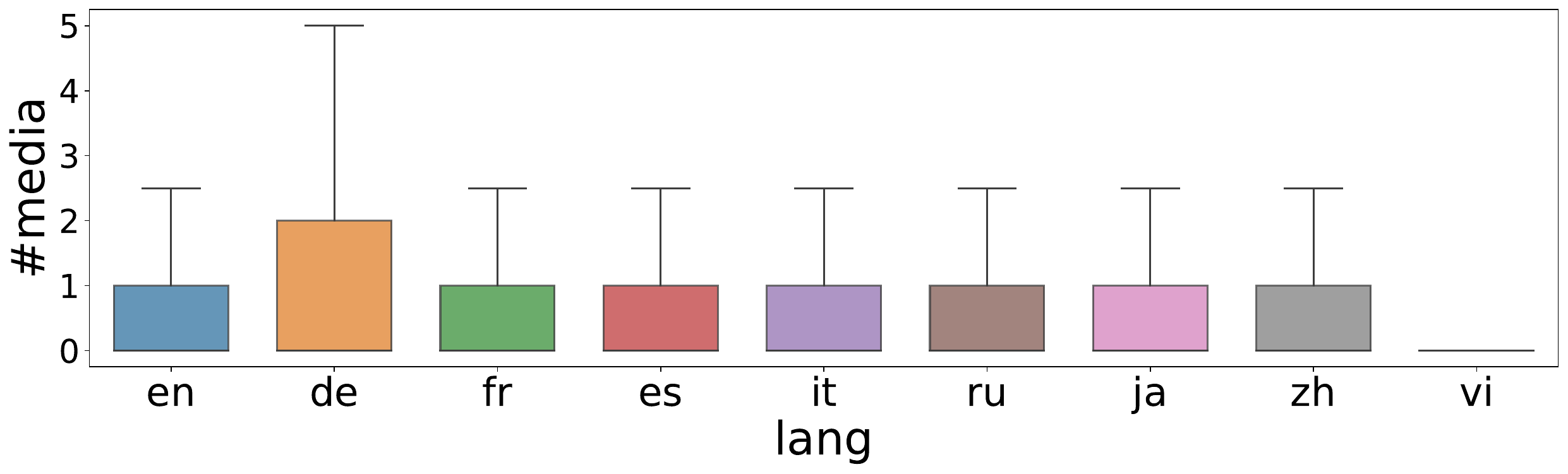}
         \caption{Media.}
         \label{fig:num_media_boxplot}
     \end{subfigure}
    \caption{Box plots of the feature distributions for the top 9 Wikipedia language versions by editing activity: English (\textit{en}), German (\textit{de}), French (\textit{fr}), Spanish (\textit{es}), Italian (\textit{it}), Russian (\textit{ru}), Japanese (\textit{ja}), Chinese (\textit{zh}) and Vietnamese (\textit{vi}). Each box plot represents the distribution of feature values of the latest revision of each article in a given language version.}
    \label{fig:features_boxplots}
\end{figure*}

To illustrate the value of the dataset, we compare the 9 largest language versions by editing activity: English, German, French, Spanish, Italian, Russian, Japanese, Chinese, and Vietnamese. Figure~\ref{fig:features_boxplots} presents box plots of the distribution of values of each feature in the latest revision of each article in these 9 versions. We observe that English Wikipedia, the largest and most popular language version, exhibits larger values in features like page length and number of references. However, the Japanese Wikipedia is the leading one in the number of sections and of wikilinks. We also note the remarkable lower values for the Vietnamese Wikipedia, a language version with a high percentage of very short articles (i.e., stubs) that were bot-generated\footnote{\url{https://stats.wikimedia.org/EN/BotActivityMatrixCreates.htm}}.

\section{Quality Modeling of Wikipedia Articles}\label{sec:qualitymodeling}

Our approach to quality modeling of Wikipedia articles across languages relies on the language-agnostic features described in the previous section. The pipeline has two stages: 1) learning feature weights, and 2) deriving pre-processing thresholds. 

In the first stage, a small sample of data is used to learn the relative weight of each of the model features (e.g., categories, text, etc). This stage is also used for testing different feature transformations such as log-normalization. In the second stage, the language-agnostic features from every article are compared against the 95th-percentile for that language edition of Wikipedia to determine what a ``high-quality'' article should attain -- e.g., if the top 5\% of articles in English Wikipedia have 14 categories, then an article with 5 categories will have a score of 0.36 ($min(1, 5/14)$) for that feature while an article with 20 categories would have a score of 1 ($min(1, 20/14)$). Certain global minimum thresholds are also set based on eye-balling the data at this stage. For example, the minimum threshold of sections is 0.1 in order to penalize bot-driven language editions of Wikipedia with many articles with lede paragraphs (i.e., 0 sections). Weights and thresholds of each feature are shown in Table~\ref{tab:weights}.

\begin{table}[h]
\begin{tabular}{l|l|p{0.5\linewidth}}
\textbf{Feature} & \textbf{Weight} & \textbf{Min. threshold for top \mbox{quality}} \\
\hline
Page length & 0.395  & 10,000 characters \\
References  & 0.181  & 0.15 ($\sim$2 references per section) \\
Sections    & 0.123  & 0.1 (1 heading at 100 chars, 2 headings at 400 chars, etc.) \\
Wikilinks   & 0.115  & 0.1 ($\sim$1 link per sentence) \\
Media       & 0.114  &  2 \\
Categories  & 0.070  & 5 \\
\end{tabular}
\caption{Weight and minimum threshold for top quality of the language-agnostic features (most language editions of Wikipedia might have thresholds higher than this minimum).}\label{tab:weights}
\end{table}

\subsection{Dataset of Predicted Quality Scores}

With our dataset of language-agnostic features of revisions of Wikipedia articles in over 300 language editions, we apply the modeling approach described above to predict their quality. Again, we store the results in CSV files with the original columns (\textit{revision\_{}id}, \textit{page\_{}id}) and a column with predicted quality score (\textit{pred\_{}qual}). In addition, we include a column with the id of each article in Wikidata (\textit{item\_{}id}) to facilitate the identification of the same Wikipedia article across language editions.

\begin{figure*}
     \centering
     \begin{subfigure}[b]{0.33\textwidth}
         \centering
         \includegraphics[width=\textwidth]{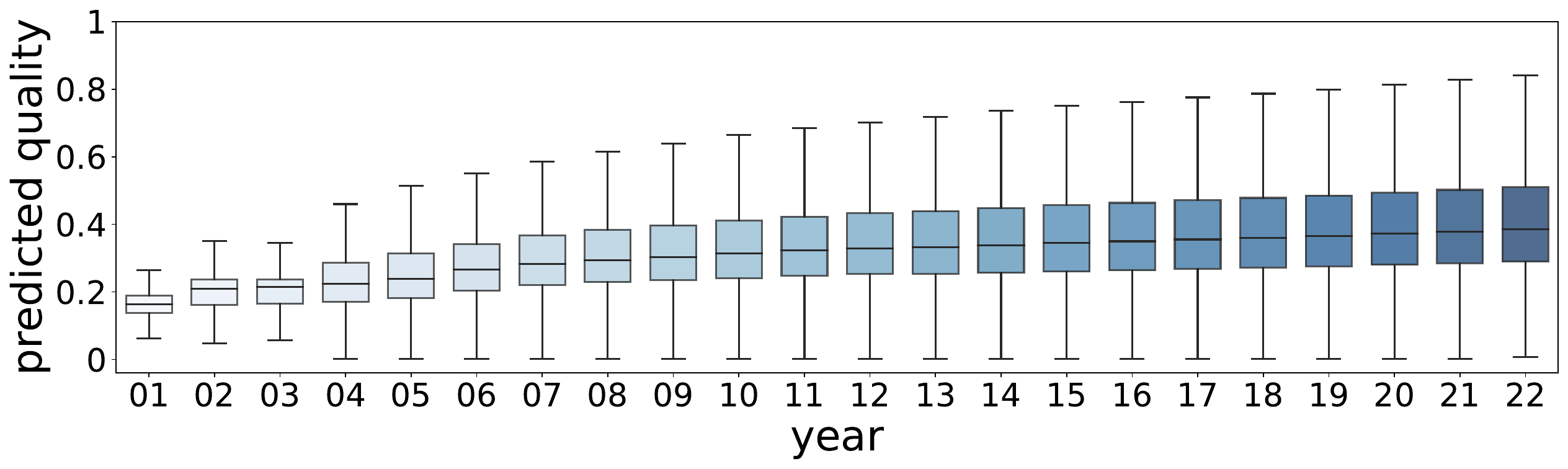}
         \caption{English Wikipedia.}
         \label{fig:enwiki_boxplot}
     \end{subfigure}
     \hfill
     \begin{subfigure}[b]{0.33\textwidth}
         \centering
         \includegraphics[width=\textwidth]{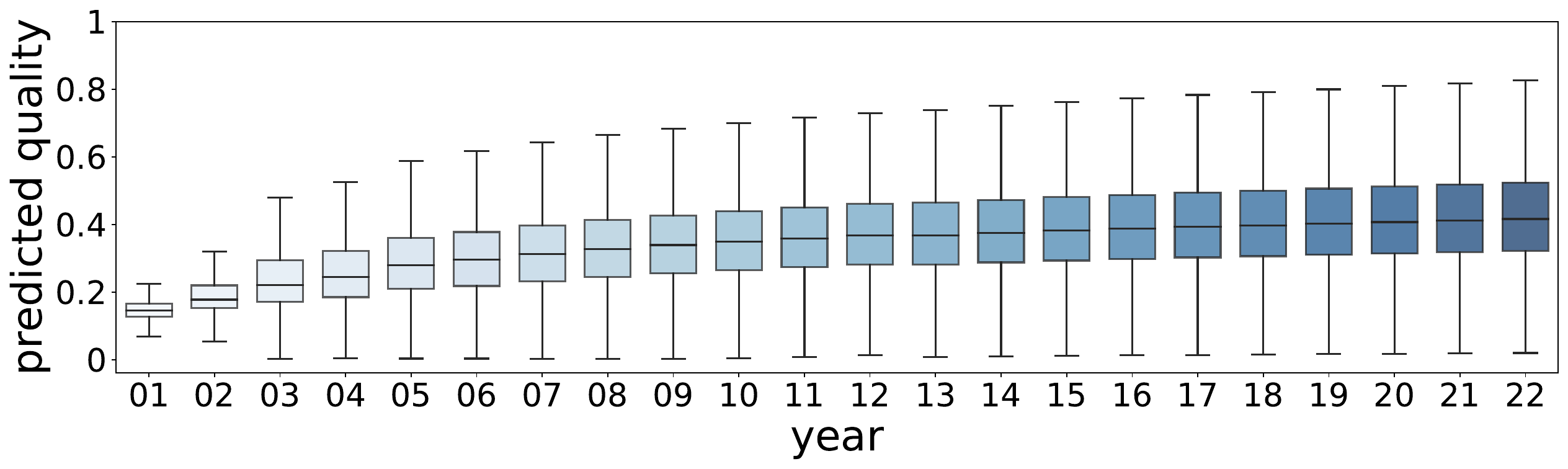}
         \caption{German Wikipedia.}
         \label{fig:dewiki_boxplot}
     \end{subfigure}
     \hfill
     \begin{subfigure}[b]{0.33\textwidth}
         \centering
         \includegraphics[width=\textwidth]{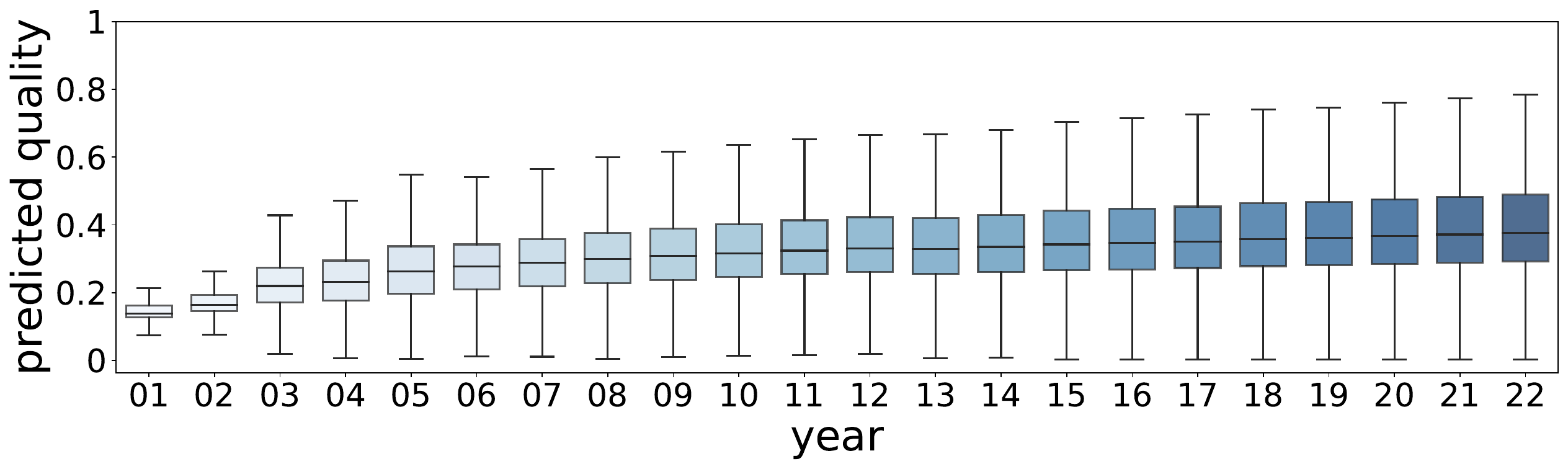}
         \caption{French Wikipedia.}
         \label{fig:frwiki_boxplot}
     \end{subfigure}
     \begin{subfigure}[b]{0.33\textwidth}
         \centering
         \includegraphics[width=\textwidth, trim={0 0 0 -2cm}, clip]{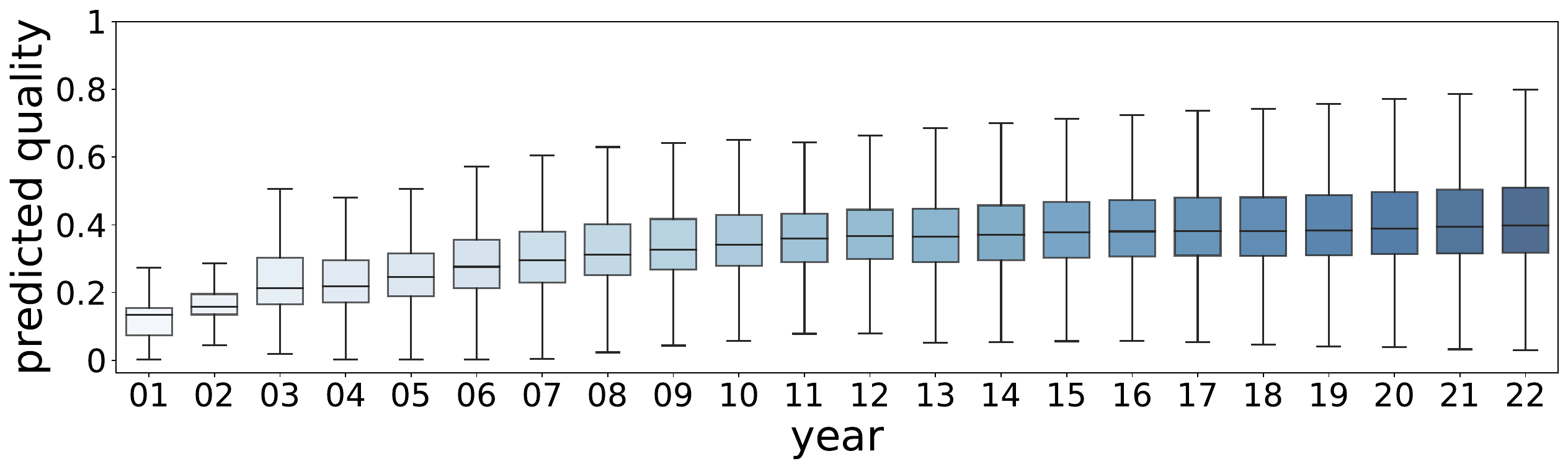}
         \caption{Spanish Wikipedia.}
         \label{fig:eswiki_boxplot}
     \end{subfigure}
     \hfill
     \begin{subfigure}[b]{0.33\textwidth}
         \centering
         \includegraphics[width=\textwidth]{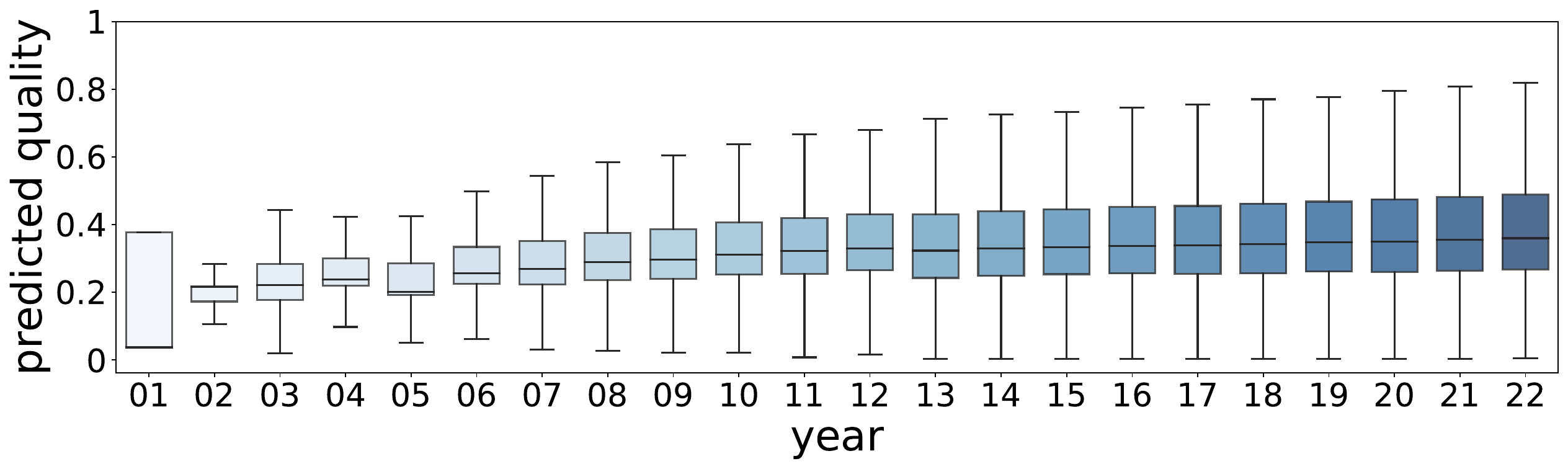}
         \caption{Italian Wikipedia.}
         \label{fig:itwiki_boxplot}
     \end{subfigure}
     \hfill
     \begin{subfigure}[b]{0.33\textwidth}
         \centering
         \includegraphics[width=\textwidth]{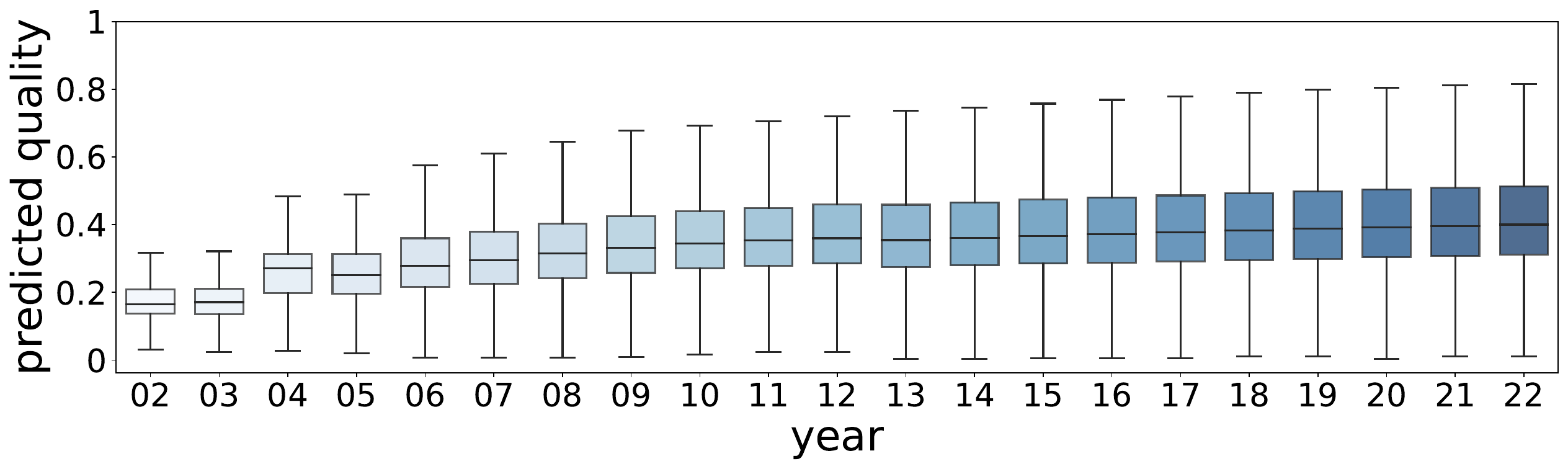}
         \caption{Russian Wikipedia.}
         \label{fig:ruwiki_boxplot}
     \end{subfigure}     
     \begin{subfigure}[b]{0.33\textwidth}
         \centering
         \includegraphics[width=\textwidth, trim={0 0 0 -2cm}, clip]{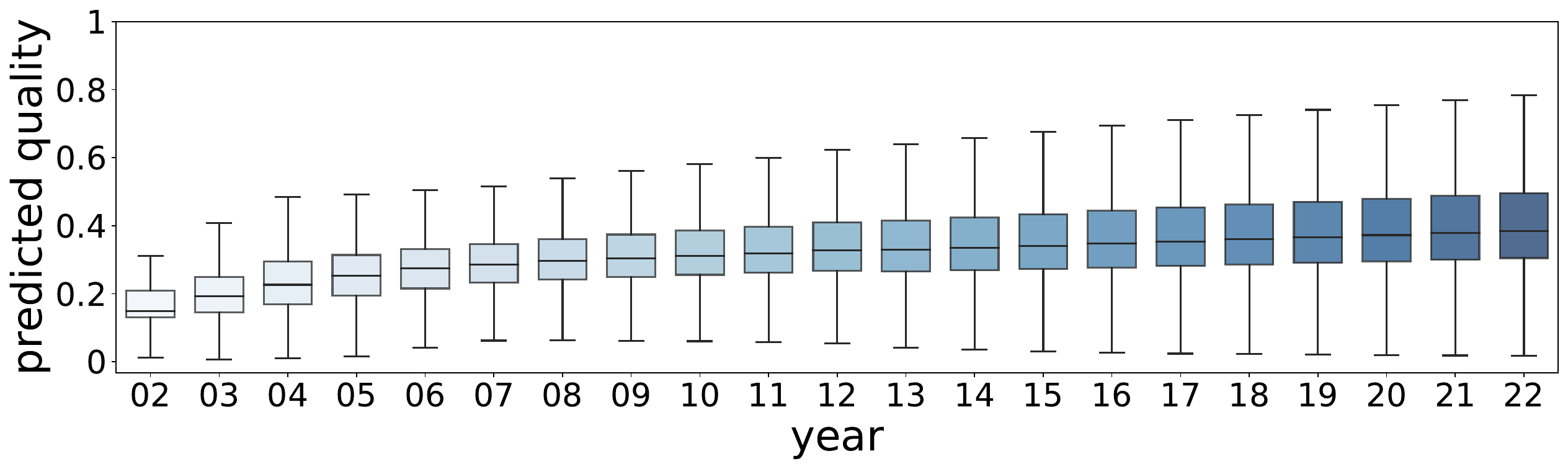}
         \caption{Japanese Wikipedia.}
         \label{fig:jawiki_boxplot}
     \end{subfigure}
     \hfill
     \begin{subfigure}[b]{0.33\textwidth}
         \centering
         \includegraphics[width=\textwidth]{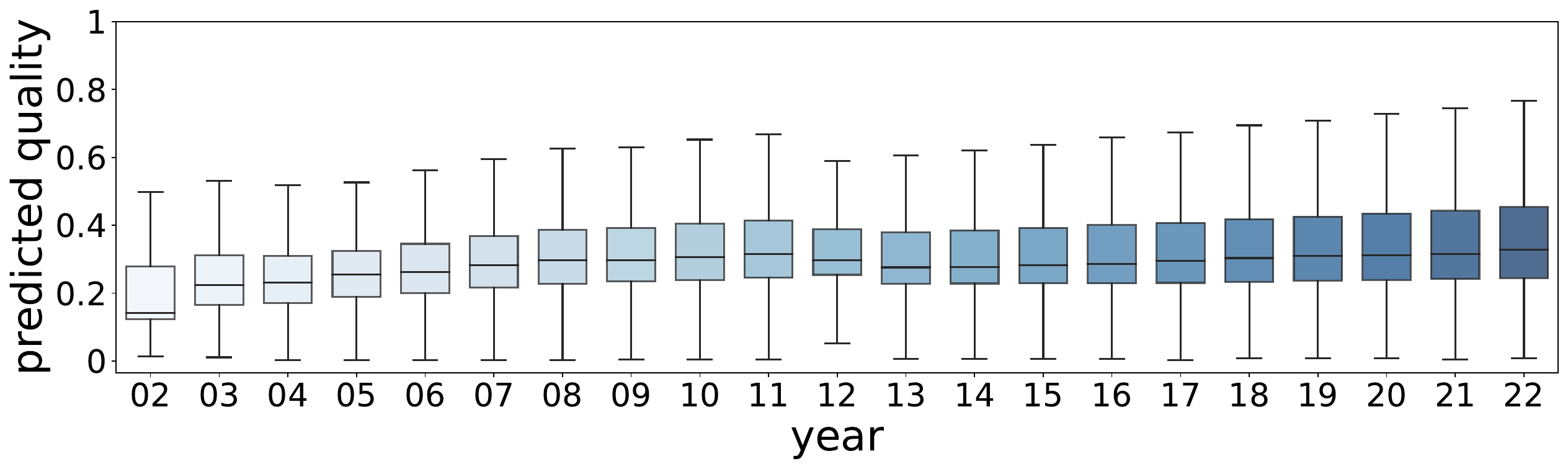}
         \caption{Chinese Wikipedia.}
         \label{fig:zhwiki_boxplot}
     \end{subfigure}
     \hfill
     \begin{subfigure}[b]{0.33\textwidth}
         \centering
         \includegraphics[width=\textwidth]{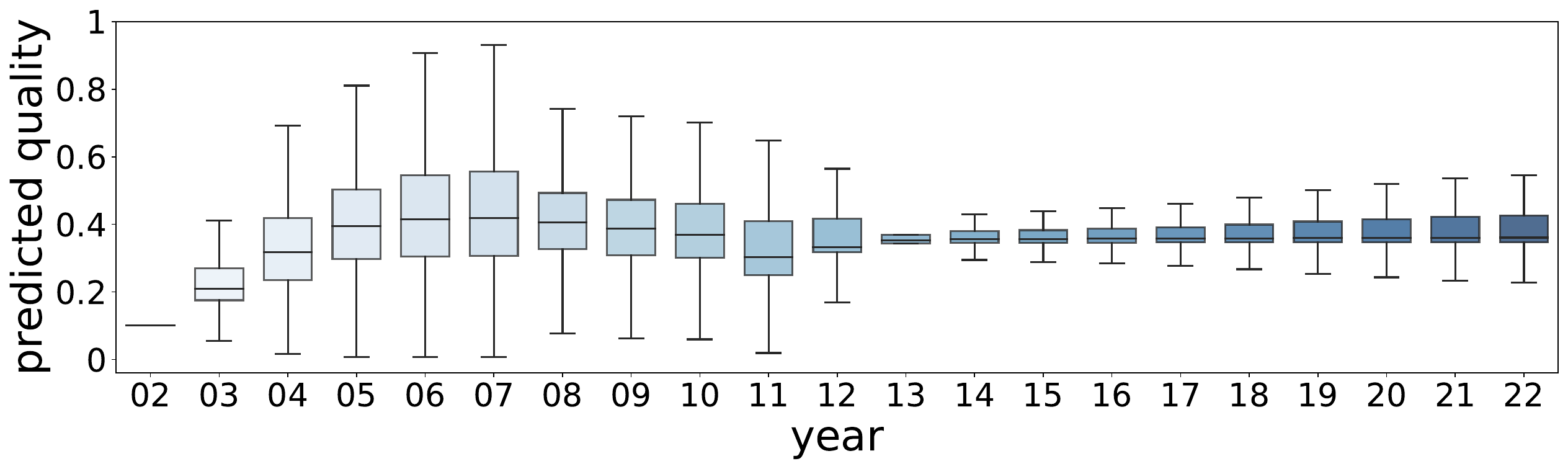}
         \caption{Vietnamese Wikipedia.}
         \label{fig:viwiki_boxplot}
     \end{subfigure}     
    \caption{Box plots of predicted article quality over time for the top 9 Wikipedia language editions by editing activity. Each box represents the predicted quality scores of the latest revision up to a given year of each article in a given language edition of Wikipedia. Color darkness corresponds to the time dimension, the darker the more recent.}
    \label{fig:quality_boxplots}
\end{figure*}

To provide a broad overview of the dataset, we analyze the evolution of article quality for the 9 largest language editions by editing activity. In particular, for each year, we select the predicted quality score of the latest revision of all existing articles until that year (included). These scores are grouped into box plots in Figure~\ref{fig:quality_boxplots}. 
For most language editions, we observe a slow but steady increase over time, with the overall quality becoming more stable in recent years. This observation can be explained by the labor of the Wikipedia editors improving the quality of articles by expanding their contents. However, this is not the case for all the language editions. Article quality in the Vietnamese Wikipedia presents a rise and fall until 2013, with values thereafter concentrated in a range more limited than in other language editions. We examined the Vietnamese language edition in the \textit{MediaWiki History} dataset\footnote{\url{https://w.wiki/7kTP}} and found an increasing ratio of bot-generated revisions until 2014 (up to 92\% of revisions in that year were written by bots) and then declining and rebounding again in the last 3 years.

\subsection{Model Evaluation}

To evaluate the effectiveness of our modeling approach by the quality assessment scheme of Wikipedia articles, we compile a set of sample test articles from both the English Wikipedia and the French Wikipedia. These test articles are used to compare the predictions made by our model against the ground-truth quality labels that have been assigned by Wikipedia editors. The term “ground truth quality” refers to the quality ratings assessed to articles by individual editors or organized groups of editors, such as WikiProject communities\footnote{\url{https://en.wikipedia.org/wiki/Wikipedia:WikiProject}}). These quality labels are added to the talk pages of the articles through a template, as mentioned in the previous section. By comparing our predictions with these ground-truth quality labels, we can assess the accuracy and reliability of our approach.
To evaluate our modeling approach to the automatic content quality assessment of Wikipedia articles, we rely on a sample of test articles from English Wikipedia and French Wikipedia. 

We extract the ground-truth quality labels in these two language editions of Wikipedia through regular expressions. We select only articles whose revision timestamp was updated before the last quality assessment appearing on their talk page. In this way, we ensure that the content of the articles has not changed substantially between the time of the ground-truth quality assessment and the revision from which we extract language-agnostic features. Thus, we select the corresponding revision of articles to create a dataset with 12,640 and 12,864 articles from English and French Wikipedia, respectively, with ground-truth quality labels. We also ensure that the dataset is composed of balanced label distributions from each of the quality classes of their respective Wikipedia language edition.

Although French Wikipedia utilizes a distinct quality scheme from that of English Wikipedia, we have established a mapping between the quality classes used in French Wikipedia and the quality assessment scheme of English Wikipedia qualitatively. For example, FA-quality articles in English Wikipedia resemble similar quality which is labeled as AdQ in French Wikipedia. The ground-truth labels extracted from French Wikipedia articles are mapped to English Wikipedia quality labels following the mapping scheme as tabulated in Table~\ref{tab:quality_labels}. This mapping helps us to compare the output of our model to a standardized quality scheme, which is considered as the ground-truth label for evaluation purposes. This way aligning the two different quality class hierarchies from two different language editions, we can effectively compare and evaluate the performance of our model based on a common quality framework.

For our testing sample of articles with ground-truth quality labels, we apply our modeling framework to compute numerical quality scores (between 0 and 1) for each of the test articles. Then, we map each output score to a quality label of English Wikipedia according to the range derived from a small sample of English Wikipedia articles (i.e., the upper limit of each quality class is the median of the predicted quality score of revisions corresponding to such class). This way the generated quality score of the test articles (i.e., here French and English articles) is mapped to the English Wikipedia quality classes.

\begin{figure} [h]
\centering
 \begin{subfigure}[b]{0.45\textwidth}
         \centering
         \includegraphics[width=\textwidth]{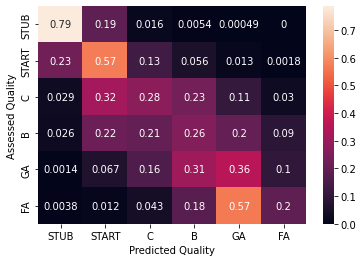}
         \caption{English Wikipedia}
         \label{fig:conf_enwiki}
     \end{subfigure}
\hfill
\begin{subfigure}[b]{0.45\textwidth}
         \centering
         \includegraphics[width=\textwidth]{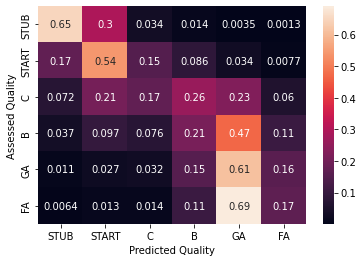}
         \caption{French Wikipedia}
         \label{fig:conf_frwiki}
     \end{subfigure}
\caption{Confusion matrix of class distributions of the articles as predicted by our model for (a) English Wikipedia and (b) French Wikipedia.}
\label{fig:conf_matrix_plots}
\end{figure}

Figure~\ref{fig:conf_matrix_plots} shows the confusion matrices for the prediction result of our model with the dataset of revisions from English and French Wikipedia. As can be seen in both matrices, the misclassification rate is lower for the quality label subgroups. For example, in the case of the English Wikipedia, articles belonging to the \textit{FA} classes are predicted as \textit{GA} to a greater extent than the other types of quality classes. This is also true for the French Wikipedia, as well as for other divisions (i.e. \textit{START/STUB}, \textit{B/C}) of quality classes. Quality classes are defined by qualitative measures only. Therefore, we conclude that distinguishing the classes of their immediate top/bottom quality is a complex task.

\begin{table}[h]
    \centering
    \begin{tabular}{c|c}
    French WP quality labels & English WP quality labels \\
    \hline
        AdQ & FA \\
        BA & GA\\
        A & B \\
        B & C \\
        BD & START \\
        ébauche & STUB \\
    \end{tabular}
    \caption{Quality classes followed in French Wikipedia and
its equivalent quality classes in English Wikipedia.}
    \label{tab:quality_labels}
\end{table}

\subsection{Model Benchmarking}

Finally, we benchmark our modeling against two baseline machine learning models. The first one is ORES, an article quality prediction framework created by the Wikimedia Foundation~\cite{halfaker2020ores}. The second one is a Random Forest (RF). We choose this second model not only to compare it to our framework but also to examine the predictive value of our set of language-agnostic features using supervised machine learning approaches. RF models are trained individually for the revisions of English Wikipedia and French Wikipedia. We select the best hyperparameter settings in the training phase. To improve the estimated performance of the RFs on all the classes, we implement 10-fold cross-validation, and the average accuracy obtained by the classifier is 0.52 for English Wikipedia and 0.51 for French Wikipedia. 

Model benchmarking against ORES and RF models is based on the following metrics:

\begin{itemize}
    \item \textbf{Spearman rank correlation (m1}). To measure the variation between ground-truth quality labels and model scores, we use Spearman's rank correlation coefficient. Ground-truth labels are converted to numerical values according to the ranks of the classes. For example, the six quality classes in English Wikipedia are \textit{FA (featured article), GA (good article), B, C, START, STUB}. Therefore, \textit{FA} labels are transformed to 1.0, \textit{GA} labels to 5/6, \textit{B} labels to 4/6, etc. Like any other correlation analysis, the coefficient is in $[-1,1]$ with 0 implying no correlation.
    
    \item \textbf{Label alignment.} 
    We quantify label alignment in three ways described below:
    
    \begin{itemize}
    
        \item \textbf{Exact match (m2)}. Percentage of predicted labels exactly assigned to the ground-truth labels. 
        
        \item \textbf{Within the same group (m3}): Percentage of predicted labels falling within the same group of ground-truth classes. Quality classes are categorized into three groups of labels: \textit{GA/FA}, \textit{C/B} and \textit{START/STUB}. 

        \item \textbf{Within one class (m4)}. Percentage of predicted labels matching within one class of ground-truth labels.       
    \end{itemize}
\end{itemize}

The results of our model benchmarking are presented in Table~\ref{tab:metric_result}. For revisions from the English Wikipedia, we observe that ORES provides better results in all metrics. 
We want to recall that ORES models incorporate language-dependent features, which might explain obtaining more accurate predictions of article quality. However, results are different in French Wikipedia, achieving slightly better predictions with RFs and our framework in certain metrics, e.g., label assignment within one class.

These results support our intuition about the noteworthy predictive value of modeling with language-agnostic features. Furthermore, the improvement using RFs -- in comparison to simpler heuristics of our framework -- suggests a promising potential of machine learning techniques with our set of language-agnostic features.

\begin{table*}[h]
    \centering
    \scalebox{1}{
    \begin{tabular}{|p{1.36cm}||p{0.6cm}|p{0.6cm}|p{0.6cm}|p{0.6cm}||p{0.6cm}|p{0.6cm}|p{0.6cm}|p{0.6cm}|| p{0.6cm}|p{0.6cm}|p{0.6cm}|p{0.6cm}|}
    \hline
    \multirow{2}{*}{\textbf{Language}} & \multicolumn{4}{c|}{\textbf{M}} & \multicolumn{4}{c|}{\textbf{ORES}} & \multicolumn{4}{c|}{\textbf{RF}} \\
    \cline{2-13}
    & m1 & m2 & m3 & m4 & m1 & m2 & m3 & m4 & m1 & m2 & m3 & m4\\
    \hline
    \textit{English} & 0.79 & 40.9 & 66.3 & 82.4 & 0.85 & 58.6 & 78.7 & 89.9 & 0.82 & 51.7 & 73.8 & 85.2 \\
    \hline
    \textit{French} & 0.76 & 40.4 & 67.9 & 83.9 & 0.79 & 50.9 & 68.4 & 83.0 & 0.80 & 51.6 & 71.9 & 83.8 \\
    \hline
    \end{tabular}}
    \caption{Benchmarking of our model (M) agains ORES and the Random Forest models (RF) for the dataset of revisions with ground-truth labels from English and French Wikipedia using the metrics: Spearman rank correlation (m1), label alignment exact match (m2), label alignment within the same group (m3), label alignment Within one class (m4).}
    \label{tab:metric_result}
\end{table*}
\section{Conclusion and Future Work}\label{sec:conclusion}

In this paper we have presented a framework to model Wikipedia article quality using language-agnostic features. Our approach transforms the unstructured and massive content of Wikipedia XML dumps into a dataset of language-agnostic features from revisions. Therefore, this resource contains a structured, smaller, and more manageable representation of the full history of all Wikipedia articles. Additionally, we have created a second dataset with the scores resulting from applying our framework that automatically assesses article quality using these features.

Our datasets have several applications. The most intuitive one is to examine the evolution of article quality in a particular language version of Wikipedia and running cross-lingual studies. For example, future research could analyze each individual feature to conduct specific studies on the number of characters, references, sections, images, and links added within and across languages, e.g., to evaluate the impact of coordinated campaigns to expand knowledge on Wikipedia~\cite{10.1145/3125433.3125475,langrock2022gender}. 

Our work complements other public datasets such as \textit{MediaWiki History}, 
\textit{Wikipedia Article Topics}~\cite{johnson2021language}, and \textit{Wiki-Reliability}~\cite{wong2021wiki}, providing a crucial missing piece. Combining all these resources with our framework and datasets would allow further research to address other relevant downstream tasks, e.g., examining content gaps in quality on a given topic, quantifying the relationship between quality and reliability of articles, measuring the impact of anonymous edits on article quality, etc.

With the release of the datasets of our work, we expect to make Wikipedia content more accessible to diverse research communities. Our language-agnostic framework aims at providing valuable input data for machine learning services in any language, including low-resourced languages in alignment with the spirit of knowledge equity. However, we should note that the evaluation of our quality assessment model was done with articles from English and French Wikipedia. While these two language editions were chosen because they allowed us to obtain comparable ``ground-truth quality'' labels, the exclusive use of these two languages for testing constitutes a limitation of this study. Therefore, future work should extend this evaluation with data from low-resource language editions of Wikipedia.

\subsection{Ethical and FAIR Considerations}\label{sec:ethics}

Our datasets are based on public data from the XML dump of the history of revisions of Wikipedia articles\footnote{\url{https://dumps.wikimedia.org/backup-index.html}}. This research was observational only and we did not use any user information from Wikipedia editors. We do not envision any potential negative societal impacts or potential misuse of this work. These released datasets conform to the FAIR principles~\cite{wilkinson2016fair} as follows:
\begin{itemize}
    \item \textbf{Findable}: The datasets have been made publicly available using the Zenodo data service that provides a permanent digital object identifier (DOI): \url{https://doi.org/10.5281/zenodo.10495081}.
    \item \textbf{Accessible}: Anyone with an Internet connection can freely access our datasets, which are licensed under CC BY-SA 4.0.
    \item \textbf{Interoperable}: The datasets are released in CSV files, a standard format for tabular data that is easily readable in most programming languages and data processing platforms.
    \item \textbf{Re-usable}: A README file is provided with metadata to facilitate the re-usability of our datasets.
\end{itemize}
Inspired by~\cite{mitchell2019model}, additional details about our framework have been documented in a model card\footnote{\url{https://meta.wikimedia.org/wiki/Machine_learning_models/Proposed/Language-agnostic_Wikipedia_article_quality}}.

\section{Acknowledgments}
We owe many thanks to Wikipedia editors, particularly those involved in assessing the quality of articles, as their work was key to building and testing our modeling approach.

\bibliography{main}

\subsection{Paper Checklist}

\begin{enumerate}

\item For most authors...
\begin{enumerate}
    \item  Would answering this research question advance science without violating social contracts, such as violating privacy norms, perpetuating unfair profiling, exacerbating the socio-economic divide, or implying disrespect to societies or cultures?
    \answerYes{Yes, this research focuses on public data not related to individuals and relies on language-agnostic features to support knowledge equity among language communities in Wikimedia.}
  \item Do your main claims in the abstract and introduction accurately reflect the paper's contributions and scope?
    \answerYes{Yes.}
   \item Do you clarify how the proposed methodological approach is appropriate for the claims made? 
    \answerYes{Yes, see details in the section describing our language-agnostic framework.}
   \item Do you clarify what are possible artifacts in the data used, given population-specific distributions?
    \answerYes{Yes, see details when describing the data and the Wikipedia quality assessment scheme.}
  \item Did you describe the limitations of your work?
    \answerYes{Yes, we compared the limitations of our approach in comparison to language-dependent approaches to quality assessment in Wikipedia.}
  \item Did you discuss any potential negative societal impacts of your work?
    \answerYes{Yes, we mentioned that we do not envision any example in this regard.}
      \item Did you discuss any potential misuse of your work?
    \answerYes{Yes, we mentioned that we do not envision any example in this regard.}
    \item Did you describe steps taken to prevent or mitigate potential negative outcomes of the research, such as data and model documentation, data anonymization, responsible release, access control, and the reproducibility of findings?
    \answerYes{Yes, these details are provided in the model card.}
  \item Have you read the ethics review guidelines and ensured that your paper conforms to them?
    \answerYes{Yes.}
\end{enumerate}

\item Additionally, if your study involves hypotheses testing...
\begin{enumerate}
  \item Did you clearly state the assumptions underlying all theoretical results?
    \answerNA{NA.}
  \item Have you provided justifications for all theoretical results?
    \answerNA{NA.}
  \item Did you discuss competing hypotheses or theories that might challenge or complement your theoretical results?
    \answerNA{NA.}
  \item Have you considered alternative mechanisms or explanations that might account for the same outcomes observed in your study?
    \answerNA{NA.}
  \item Did you address potential biases or limitations in your theoretical framework?
    \answerNA{NA.}
  \item Have you related your theoretical results to the existing literature in social science?
    \answerNA{NA.}
  \item Did you discuss the implications of your theoretical results for policy, practice, or further research in the social science domain?
    \answerNA{NA.}
\end{enumerate}

\item Additionally, if you are including theoretical proofs...
\begin{enumerate}
  \item Did you state the full set of assumptions of all theoretical results?
    \answerNA{NA.}
	\item Did you include complete proofs of all theoretical results?
    \answerNA{NA.}
\end{enumerate}

\item Additionally, if you ran machine learning experiments...
\begin{enumerate}
  \item Did you include the code, data, and instructions needed to reproduce the main experimental results (either in the supplemental material or as a URL)?
    \answerYes{Yes, we provide all the details in the model card, including a link to the code to reproduce the model evaluation.}
  \item Did you specify all the training details (e.g., data splits, hyperparameters, how they were chosen)?
    \answerYes{Yes, we provide all the details in the model card.}
     \item Did you report error bars (e.g., with respect to the random seed after running experiments multiple times)?
    \answerNo{No, as machine learning models are only used for benchmarking purposes (our framework relies on a heuristic approach).}
	\item Did you include the total amount of compute and the type of resources used (e.g., type of GPUs, internal cluster, or cloud provider)?
    \answerNo{No, as machine learning models are only used for benchmarking purposes (our framework relies on a heuristic approach).}
     \item Do you justify how the proposed evaluation is sufficient and appropriate to the claims made? 
    \answerYes{Yes, see details in the subsection describing our model benchmarking.}
     \item Do you discuss what is ``the cost`` of misclassification and fault (in)tolerance?
    \answerYes{Yes, see details in the subsection describing our model benchmarking.}
  
\end{enumerate}

\item Additionally, if you are using existing assets (e.g., code, data, models) or curating/releasing new assets, \textbf{without compromising anonymity}...
\begin{enumerate}
  \item If your work uses existing assets, did you cite the creators?
    \answerYes{Yes.}
  \item Did you mention the license of the assets?
    \answerYes{Yes.}
  \item Did you include any new assets in the supplemental material or as a URL?
    \answerYes{Yes, we provide references to resources like the datasets in Zenodo, and the model card.}
  \item Did you discuss whether and how consent was obtained from people whose data you're using/curating?
    \answerYes{Yes, our datasets are based on public data from the XML dump of the history of revisions of Wikipedia articles.}
  \item Did you discuss whether the data you are using/curating contains personally identifiable information or offensive content?
    \answerYes{Yes, our datasets does not contain personally identifiable information or offensive content.}
\item If you are curating or releasing new datasets, did you discuss how you intend to make your datasets FAIR (see \citet{wilkinson2016fair})?
\answerYes{Yes.}
\item If you are curating or releasing new datasets, did you create a Datasheet for the Dataset (see \citet{gebru2021datasheets})? 
\answerNo{No, but, inspired by~\cite{mitchell2019model}, a model card was created.}
\end{enumerate}

\item Additionally, if you used crowdsourcing or conducted research with human subjects, \textbf{without compromising anonymity}...
\begin{enumerate}
  \item Did you include the full text of instructions given to participants and screenshots?
    \answerNA{NA.}
  \item Did you describe any potential participant risks, with mentions of Institutional Review Board (IRB) approvals?
    \answerNA{NA.}
  \item Did you include the estimated hourly wage paid to participants and the total amount spent on participant compensation?
    \answerNA{NA.}
   \item Did you discuss how data is stored, shared, and deidentified?
   \answerNA{NA.}
\end{enumerate}

\end{enumerate}

\end{document}